\documentstyle[12pt]{article}

\font\frak=eufm10 scaled\magstep1

\font\black=msbm10 scaled\magstep1
\font\bigblack=msbm10 scaled\magstep 2
\font\bbigblack=msbm10 scaled\magstep3

\def\goth #1{\hbox{{\frak #1}}}

\def\bigfield #1{\hbox{{\bigblack #1}}}
\def\bbigfield #1{\hbox{{\bbigblack #1}}}
\def\Bbb #1{\hbox{{\black #1}}}
\def\v #1{\vert #1\vert}             
 
\def\m #1 #2{(-1)^{{\v #1} {\v #2}}} 
\def\pd#1#2{\frac{\partial#1}{\partial#2}}
\def\<#1>{\langle#1\rangle}        
\def\>#1{{\bf #1}}                
\def\f(#1,#2){\frac{#1}{#2}}

\def\dt2#1{\frac{d^2 #1}{dt^2}}

\def\G{{\Gamma}}
\def\w{\omega}                     
	     
\def\X{{{\goth X}}}                     
\def\big R{{\hbox{{\bigfield R}}}}
\def\bbig R{{\hbox{{\bbigfield R}}}}

\def\dim{\hbox{{\rm dim}}}                                                                      
 
\def\be{\begin{equation}}
\def\ee{\end{equation}}
\def\bea{\begin{eqnarray}}
\def\eea{\end{eqnarray}}
\def\dim{\hbox{{\rm dim}}}

\begin{document}

\centerline{\bf Presymplectic Geometry and Fermat's} 
 \medskip
\centerline{\bf Principle for anisotropic media} 

\vskip 2cm

\centerline{ {\sc J.F. Cari\~nena*,  
  and  J. Nasarre ${}^{\mbox{\small{\S\S}}}$ 
}}
\vskip 1cm

\centerline{*Depto. F\'{\i}sica Te\'orica, Univ. de Zaragoza, 50009 
Zaragoza,
Spain.}

 \centerline{\S\S  Seminario~de Matem\'aticas, IES Miguel Catal\'an,
 50009 Zaragoza,  Spain. }

\vskip 2cm

\centerline{\sc Abstract}
{\small The tools of presymplectic geometry are used to study  
light rays trajectories  
in anisotropic media.
 }

\vskip 1cm

The usefulness of the Lie--Hamilton optics in many different problems as ray tracing 
for ray design and computation of aberrations suggests the study of what happens 
for anisotropic media, because of the recent interest in the use of anisotropic
optical material. This motivated a very recent paper {\cite{Wolfuno}}
where the Hamiltonian
formulation of geometric anisotropic optics was studied. The theory was reexamined
in \cite {Torres}. The basic principle of the theory is the celebrated Fermat's principle
of least time (or extremal time if reflection is also allowed). In other 
words, light 
rays connecting points $A$ and $B$ are lines in the space in such a way that they satisfy   
the following variational condition:
\begin{equation}
\delta \int ^B_A n\, ds=0.  \label{Fermat}
\end{equation}

The refractive index of the medium is given by the quotient $n=\frac cv$, 
and then physics tell us that $n>1$. 

In a recent paper \cite{nuestro} we analysed from a geometric perspective  
the relationship for the case of isotropic media  of the problem of 
determination of extremal curves for  (\ref{Fermat})
with that of geodesics of a Riemannian metric conformal to the 
Euclidean metric.
We will consider here  the case
in which the medium is not isotropic but it may depend on the velocity,
or more specifically, on the direction of 
the ray. In this last case the problem cannot be  reduced to a 
problem of Riemannian geometry as it happened when the refractive index 
$n$ only depended on the position. So, the techniques of Presymplectic
geometry are unavoidable for dealing with this dependence of the refractive
index with the ray direction.

To begin we remark the strong similarity of Fermat's principle with the more 
traditional   Hamilton's principle of Classical Mechanics, 
with a Lagrangian function 
 given by 
\be
L=n\, \sqrt {g(v,v)}.  \label{Lopt}
\ee
This Lagrangian function is differentiable only in the set of velocity
phase space obtained by removing the null velocity points, i.e.,
the zero section of the tangent bundle. Moreover, the Lagrangian $L$
is homogeneus of degree one,
\be 
v^i\pd L{v^i}=L,
\ee
and consequently the corresponding energy function vanishes identically. 
Therefore the Lagrangian is singular, because taking derivatives with 
respect to $v^j$ of both sides of the preceding equation we obtain  
\be
\pd{^2L}{v^i\partial v^j}\, v^i=0,
\ee
and then the Hessian matrix 
\be 
W_{ij}=\pd{^2L}{v^i\partial v^j}\label{hessian}
\ee
is singular.  
The theory should be carefully reexamined 
using the tools of Presymplectic Geometry, as it was done in \cite{nuestro}
for the isotropic media, where  the tools of Riemannian Geometry were 
shown to be very useful. Actually, it is possible to show that in the 
latter case the solution curves 
 for  the regular Lagrangian system described by the regular 
Lagrangian 
\be
\Bbb L=\frac 12 n^2 \,g(v,v) 
\ee
are just the curves solution of the original 
problem,
even if the curves are reparametrized (see e.g. 
\cite{polacos} and references therein). Our aim is to analyse what
happens in the more general case in which the refractive index can depend
on the ray direction, i.e., the refractive index is a homogeneous function
of degree zero of velocities in the set obtained by removing the 
zero section of the velocity phase espace.

The geometric approach to Lagrangian Classical mechanics uses as 
velocity phase space the tangent bundle $TM$ of  the configuration space
$M$ that is assumed to be a differentiable manifold of dimension $N$.
>From now on we will follow the notation used in
\cite {Ranada}.
The tangent  structure is characterized by
 a $(1,1)$--tensor field called vertical endomorphism $S$ that in 
terms of natural 
  coordinates $(q^i,v^i)$ of the tangent bundle $TM$ 
	is given by
\be     
S = \pd{}{v^i }\otimes   dq^i.  
\ee

Given a   
function $L\in C^\infty(TM)$, we may define an exact 2--form in $TM$,
$\w_L=-d\theta_L  $, with the 1--form $\theta_L$ being defined by 
$\theta_L  =dL\circ S$, and a function $E_L=
\Delta (L)-L$, called energy function. In the above mentioned coordinates of 
$\,TM$ we have the following expressions:
\begin{eqnarray}     \theta_L  & =& \pd{L}{v^i }\,dq^i,   \\
      \w_L &=& \pd{^2L}{q^i\partial v^j}\, dq^j\wedge dq^i + 
\pd{^2L}{v^i\partial v^j}\,dq^i\wedge dv^j,     \\
\Delta&=& v^i \pd{}{v^i } \\
 E_L &=& v^i \pd{L}{v^i } - L,    
 \end{eqnarray}
Here $\Delta \in \X(TM)$ denotes the Liouville vector field generating
dilations along the fibres. When $\omega_L$ is nondegenerate, i.e., 
the Hessian matrix (\ref{hessian}) is regular, it defines a  
symplectic structure on $TM$, and a vector field $\Gamma_L$  uniquely 
determined by 
$i(\Gamma_L)\omega_L=dE_L$. 
 
 We will next examine the problem of light rays, 
 even for the more general case in which 
the refractive index $n$ depends on the ray direction. This means that 
the refractive  index  must be homogeneous of degree zero in the
velocities, $\Delta n=0$. We define a new Lagrangian $\Bbb L=\frac 12 L^2$. 
The Lagrangian $\Bbb L$ is homogeneous of degree two in the velocities,
  $\Delta {\Bbb L}=2{\Bbb L}=L^2$, and then $E_{\Bbb L}={\Bbb L}$.   

Now taking into account that $\theta_{\Bbb L}=d\Bbb L\circ S$ we see that 
the  Liouville 1--form  $\theta_{\Bbb L} $ is proportional to $\theta_L$,
namely,
 $\theta_{\Bbb L} =L \,\theta_ L $, (see \cite {Lopez}) and then  
\be
\w_L =\frac {1} {L}\,\w_{\Bbb L} +
  \frac {1} {L^3}\,d\Bbb L\land{\theta}_{\Bbb L}.  
\ee

 As indicated above, when the refractive index $n$ does not depend on 
 velocities the 
  2--form $\w_{\Bbb L}$ is regular. In this more general case, however, it may be
 singular, because 
\be
\w_{\Bbb L}=L\,\w_L-dL\wedge \theta _L
\ee
means that 
\be
\w_{\Bbb L}\,^{\wedge N}=L^N\,\w_L\,^{\wedge N}-N \,L^{N-1}\,dL\wedge 
\theta_L\wedge 
\w_L\,^{\wedge (N-1)}=-N \,L^{N-1}\,dL\wedge \theta_L\wedge  
\w_L\,^{\wedge (N-1)},
\ee
and then  $\w_{\Bbb L}\,^{\wedge N}$ can be identically null and in this case 
$ {\Bbb L}$ would be singular.

In the following we will restrict ourselves to the case in which ${\Bbb L}$
is regular and therefore there will be a uniquely defined
vector field 
   $\Gamma_{ \Bbb L}$   
such that 
\be
i(\Gamma_{\Bbb L}) \w_{\Bbb L}=dE_{\Bbb L}=d{\Bbb L},
\ee
and then
\be
i(\G_{\Bbb L})(d{\Bbb L})=\G_{\Bbb L} ({\Bbb L})=0.   
\ee
Moreover, $\G_{\Bbb L}$ is a second order differential equation vector 
field.

First we check that the Liouville vector field $\Delta$ lies in 
$\ker \w_L$. Indeed,
\be
 i(\Delta) \w_{L} =
 \frac {1} {L} i(\Delta) \w_{\Bbb L} + 
 \frac {1} {L^3}\,\Delta \Bbb L \,\theta_{\Bbb L}
- \frac {1} {L^3} \,d{\Bbb L} \,(i(\Delta) \theta_{\Bbb L}),
\ee
and taking into account that $\Delta$ is vertical and $\theta_{\Bbb L}$
semibasic, the last term vanishes. Moreover,
\be
i(\Delta) \w_{\Bbb L} = - {\cal L}_{\Delta}\theta_{\Bbb L} +
d(i(\Delta)\theta_{\Bbb L}) = - {\cal L}_{\Delta}\theta_{\Bbb L} =
-\theta_{\Bbb L}
\ee
and therefore, taking into account that $\Delta \Bbb L = 2\Bbb L = L^2$
we find that
\be
i(\Delta) \w_L = 0.
\ee
Secondly, ${\G}_{\Bbb L}$ is also in the kernel of $\w_L$, because
\be
i(\G_{\Bbb L})\w_L =
\frac {1} {L} i(\G_{\Bbb L}) \w_{\Bbb L} + 
 \frac {1} {L^3}(\G_{\Bbb L} \Bbb L)\,
\theta_{\Bbb L} - \frac {1} {L^3}\, d{\Bbb L} \,[i(\G_{\Bbb L})\theta_{\Bbb L}]
\ee
and $\G_L$ being a SODE,
\be
i(\G_{\Bbb L})\theta_{\Bbb L}=(d\Bbb L\circ S)(\G_{\Bbb L}) =
\Delta \Bbb L = 2\Bbb L =L^2
\ee
and therefore
\be
i(\G_{\Bbb L})\w_L = 0.
\ee

 Finally under the assumption that $\Bbb L $ is regular, $ \ker\w_L $ 
is generated 
by $\Delta$ and $\G_{\Bbb L}$. Indeed, given a vertical vector field
$V\in \ker \w_L$, then,
\be
0= i(V)\w_L = \frac {1} {L} i(V) \w_{\Bbb L} + 
 \frac {1} {L^3}\,V({\Bbb L})\, \theta_{\Bbb L}
\ee
and in particular
\be
\frac {1} {L} i(\Delta) \w_{\Bbb L} = 
 -\frac {1} {L^3}\,\Delta ({\Bbb L})\, \theta_{\Bbb L}
=-\frac {1} {L} \theta_{\Bbb L}
\ee
and therefore
\be
 i(V) \w_{\Bbb L}= \frac {V(\Bbb L)} {L^2} \,i(\Delta) \w_{\Bbb L}
\ee
and as $\w_{\Bbb L}$ is assumed to be regular, $V$ should be proportional
to $\Delta$. Then, $\dim V(\ker\w_L) = 1$ and because of the relation 
$\dim  (\ker\w_L)\le 2\,\dim  V(\ker\w_L)$ (see \cite {Ibort}), we can 
conclude 
that $\dim  (\ker\w_L)=2$. 

Under these circumstances is possible to apply 
the reduction theory  of presymplectic manifolds, following
the ideas developed by Marsden and Weinstein  \cite{Marsden}.
Presymplectic  structures may arise either when using some constants of motion
for reducing the phase space or also when the Lagrangian that has been  
chosen is singular. Then we will have a pair  $(P_0,\Omega_0)$ where $\Omega_0$
is a closed but degenerate 2--form. A consistent solution of the dynamical equation
can only be found in some points, leading in this way to the  final 
constraint submanifold $P$
 introduced by Dirac (see e.g. \cite{Pepindos}). 
 The pull back  $\Omega$ of the form $\Omega_0$ on this manifold will 
 be assumed to be of constant rank.
The recipe for dealing with these systems was given by Marsden and Weinstein
\cite{Marsden}.
First, in every point $m\in P$, $\ker \Omega_m$ is a $k$--dimensional linear 
space, so defining what is called a $k$--dimensional distribution. The important
point is that closedness of $\Omega$ is enough to warrant that the distribution 
is integrable (and then it is called foliation):  for any 
point $m\in P$, there is a $k$--dimensional submanifold of $P$ passing 
through $m$
and such that the tangent space at any point $m'$ of this surface 
coincides with $\ker \Omega_{m'}$. Such integral $k$--dimensional submanifolds
give a foliation of $P$ by disjoint leaves and in the  case in which the quotient
space $\widetilde P=P/\ker \Omega$ is a differentiable manifold, then it 
is possible to define a nondegenerate closed 2--form  $\widetilde \Omega $
in $\widetilde P$ such that $\widetilde \pi^*\widetilde \Omega=\Omega$.
Here $\widetilde \pi:P\to \widetilde P$ is the natural projection. 
It suffices to define $\widetilde \Omega(\widetilde v_1, \widetilde v_2)=
\Omega(v_1,v_2)$, where $v_1$ and $v_2$ are tangent vectors to $P$ 
projecting under $\widetilde \pi_*$ onto 
$\widetilde v_1$ and $\widetilde v_2$ respectively. The symplectic
space $(\widetilde P,\widetilde\Omega)$ is said to be the reduced space.
We will illustrate  the method finding 
coordinates adapted to the distribution defined by the kernel $\ker \w_L$ 
of the presymplectic structure defined by the singular optical Lagrangian 
in the case of a system in which either the index $n$   
depends on the third coordinate $x^3$ alone or  the very interesting 
case in which the system is anisotropic and $n$ is a function of 
the ray direction. We will determine the quotient reduced space and we will look for 
Darboux coordinates in this reduced symplectic manifold.
Once Darboux coordinates have been found we can consider 
the problem from the active viewpoint and take advantage of the 
algebraic methods recently developed for computing aberrations 
(see e.g. \cite {Dragt}). 

 Let us now consider the most general isotropic case in which the
refractive 
index of the medium is not constant but it is given by a smooth 
function $n(x^1,x^2,x^3)$.
Fermat's principle suggests us to consider the corresponding mechanical problem 
described by a singular Lagrangian $L(q,v)=[g(v,v)]^{1/2}$, where $g$ is
 a metric conformal to the Euclidean metric $g_0$,
\be
g(v,w)= n^2 g_0(v,w).
\ee

This problem was analysed in \cite{Lopez} where, as above is cited,
 it was shown that 
its study can be reduced to that of a regular Lagrangian 
$\Bbb L=\frac 12 L^2$. This
Lagrangian $\Bbb L$ is quadratic in velocities and the dynamical 
vector field $\G_{\Bbb L}$ solution of the dynamical equation
$i(\G_{\Bbb L})\w_{\Bbb L}=
dE_{\Bbb L}=d{\Bbb L}$ is not only a second order differential 
equation vector field
 but, moreover, it is a spray  \cite{CP}, 
 the projection onto
 $\Bbb R^3$ of its integral curves being the geodesics of the Levi--Civita 
connection defined by $g$. Then, $\G_{\Bbb L}$ is
the geodesic spray given by
\be
\G_{\Bbb L}=v^i\pd {}{q^i}-\G^i\,_{jk}v^jv^k\pd {}{v^i} ,
\ee
where the Christoffel symbols $\G^i\,_{jk}$ are
\be
\G^i\,_{jk}=\frac 12 g^{il}\left[ \pd{g_{kl}}{q^j}+\pd{g_{jl}}{q^k}
-\pd{g_{jk}}{q^l}
\right]
\ee
 with $g^{ij}$ being the inverse matrix of $g_{ij}$.

In the particular case we are considering where $g(v,w)= n^2 g_0(v,w)$,
it was also shown above that the kernel of $\w_L$ is two--dimensional 
and it is generated by  $\G_{\Bbb L}$  and the Liouville vector field $\Delta$.
The distribution defined by $\ker \w_L$ is integrable because
$\w_L$ is closed;
actually $[\Delta,  \G_{\Bbb L}]= \G_{\Bbb L}$ and the distribution is also 
generated by $\Delta$ and 
$K$ defined by $K=\frac 1{v^3}  \G_{\Bbb L}$, for which $[\Delta, K]=0$. 
In cartesian coordinates the Christoffel symbols are expressed as follows:
\be
\G^i\,_{jk}=\frac 1{n}\left[ \pd n{x^j}\,\delta^i_k+\pd n {x^k}\,\delta^i_j-
\pd n{x^i}\,\delta^j_{k}
\right].
\ee
and the vector field $K$ is given as
\be
K=\frac 1{v^3}\left[v^i\pd{}{x^i}- \left(\frac 2n v^i( v.\nabla n)-
 \frac{\| v \|^2}{n}\pd n{x^i}\right)\pd {}{v^i}\right].
\ee

 The theory of distributions suggests us the introduction of new local
 coordinates
$y^i=F^i(x,v)$, $i=1,\ldots ,6$, adapted to the distribution defined
by  $\ker \w_L$,
i.e., such that $K=\pd {}{y^3}$, $\Delta =\pd {}{y^6}$ (see \cite{Cr83}). 
The search for these new
coordinates is based on the solution of
the partial differential equation system
$$KF^1=1, \quad \Delta F^1=0,\quad KF^2=0, \quad \Delta F^2=1,$$
and 
$$ 
KF^{2+a}=0, \quad \Delta F^{2+a}=0, \, {\rm{for}} a=1\ldots, 4.
$$

The explicit computation of these functions depends very much on
the choice of the function $n(x^1,x^2,x^3)$. We will illustrate next
the theory with an particular example. If $n$ only depends on $x^3$, 
the presymplectic form can be written in the way 
\begin{eqnarray}
\w_L &=& dx^1\wedge d\left(\frac {nv^1}{\sqrt {{v^1}^2+{v^2}^2+{v^3}^2}}\right)
+dx^2\wedge d\left(\frac {nv^2}{\sqrt {{v^1}^2+{v^2}^2+{v^3}^2}}\right)
\\ &+& \frac {n{v^3}^2}{\left({v^1}^2+{v^2}^2+{v^3}^2\right)^{3/2}}
\left[v^1d\left(
\frac {v^1}{v^3}\right)\wedge dx^3+
v^2d\left(
\frac {v^2}{v^3}\right)\wedge dx^3\right],
\end{eqnarray}
and the dynamical vector field is
\begin{eqnarray} 
\G_{\Bbb L} &=& v^i\pd {}{x^i}-\frac 2n v^1v^3\frac {dn}{dx^3}\pd{}{v^1}-
\frac 2n v^2v^3\frac {dn}{dx^3}\pd{}{v^2} \\ &+&
\frac 1n ({v^1}^2+{v^2}^2-{v^3}^2)
\frac {dn}{dx^3}\pd{}{v^3}.
\end{eqnarray}

After some calculations we find the solution 
of the former systems (see \cite{nuestro}).
According to this, we will do the following choice for the new coordinates:
\begin{eqnarray} y^1 &=& x^1-\frac {v^1}{v^2} \, x^2,\\ 
 y^2 &=& x^2-\int _0^{x^3}
\frac {C_3}{\sqrt{(n^2(\zeta)-C_3^2)(1+C^2_1)}}\,d\zeta,\\  y^3 &=& x^3\\
y^4 &=& \frac {nv^1}{\sqrt{ {v^1}^2+{v^2}^2+{v^3}^2}},\\ 
y^5 &=& \frac {nv^2}{\sqrt{{v^1}^2+{v^2}^2+{v^3}^2}},\\ 
y^6 &=& \log \left[n\sqrt {{v^1}^2+{v^2}^2+{v^3}^2}\,\right],
\end{eqnarray}
where 
\be
C_1=\frac {v^1} {v^2}
\ee
and
\be
C_3= n\sqrt{\frac{{v^1}^2+{v^2}^2} {{v^1}^2+{v^2}^2+{v^3}^2}}, 
\ee
doing the inverse change and after some easy calculations remains
\be
\widetilde\w_L=d\left(y^1+\frac {y^4}{y^5}\,y^2\right)\wedge dy^4
+dy^2\wedge dy^5,
\ee
which shows that 
\begin{eqnarray} \xi^1 &=& y^1+\frac {y^4}{y^5}y^2=x^1-\frac {v^1}{v^2}
\int_0^{x^3}\frac {C_3}{\sqrt{(n^2(\zeta)-C_3^2)
(1+C^2_1)}}\,d\zeta,\\ \xi^2 &=& x^2-\int_0^{x^3}\frac {C_3}
{\sqrt{(n^2(\zeta)-C_3^2)
(1+C^2_1)}}\,d\zeta
\end{eqnarray}
and the corresponding 
\be
\eta^1=y^4=\frac {nv^1}{\sqrt {{v^1}^2+{v^2}^2+{v^3}^2}},\quad 
\eta^2=y^5=\frac {nv^2}{\sqrt {{v^1}^2+{v^2}^2+{v^3}^2}}
\ee
are Darboux coordinates for the symplectic form induced in the quotient
space.

Let us now consider the 
 particular but important case  case in which  the refractive index becomes
 constant out of  a region. If for $x^3>L$, the index $n$ is constant,
 the above mentioned Darboux coordinates $\xi^1$ and $\xi^2$ are
$$
\xi^1=x^1-\frac {v^1}{v^2} \frac {C_3x^3}{\sqrt{(n^2-C_3^2)
(1+C^2_1)}},\quad \xi^2=x^2- \frac {C_3x^3}{\sqrt{(n^2-C_3^2)
(1+C^2_1)}},
$$
up to a constant, and from the expresions of $C_1$ and $C_3$ 
we see  that the Darboux coordinates become 
\be
x^1-\frac {v^1}{v^3}x^3,\quad x^2-\frac {v^2}{v^3}x^3, \quad \frac {nv^1}
{\sqrt{{v^1}^2+{v^2}^2+{v^3}^2}},\quad  \frac {nv^2}
{\sqrt{{v^1}^2+{v^2}^2+{v^3}^2}},
\ee
in full agreement with \cite {Wolfdos}. Therefore, for an optical system such that
the refractive index depends only on $x^3$ and, furthermore, 
 the region in which the index is not constant is bounded, we can choose
Darboux coordinates by fixing a $x^3$ outside this region and taking Darboux 
coordinates for the corresponding problem of constant index. This justify
the choice of coordinates as usually done for the  ingoing and outgoing light rays in the
corresponding constant index media, i.e. it shows the convenience of using flat screens
in far enough regions on the left and right respectively, and then this 
change of Darboux 
coordinates seems to be, from an active viewpoint, 
a canonical transformation (see \cite{Gu84}).
    
We will next find the symplectic structure arising in an 
anisotropic optical medium, as well as some Darboux coordinates for it. 
We recall what we are only considering anisotropic media for which the
  refractive index
depends only on the ray direction, i.e., $n=n(v)$ but $\Delta n=0$. In this case
the presymplectic form remains as 
\begin{eqnarray}
\w_L &=& \left[ \| v \| \frac {{\partial}^2 n}
 {{\partial}x^j{\partial}v_i} +
\frac {v_i} {\| v \|}\frac {{\partial} n} {{\partial}x^j}
\right] dx^{i}\land dx^j
\\ &+& \left[ \| v \| \frac {{\partial}^2 n}
 {{\partial}v_j{\partial}v_i} +
\frac {v_i} {\| v \| }\frac {{\partial} n} {{\partial}v_j}+
\frac {v_j} {\| v \| }\frac {{\partial} n} {{\partial}v_i}
-\frac {n} {\| v \|^3 }v_iv_j
+\frac {n} {\| v \|}{\delta}^{i}_j\right] dx^{i}\land dv_j , 
\end{eqnarray}  
and the  vector field associated with $\Bbb L$
\be
\Gamma_{\Bbb L}= v_i\frac {\partial} {{\partial}x^i}.
\ee
 We still have that
\be
\left[\Delta ,\G_{\Bbb L}\right]=\G_{\Bbb L}, 
\ee
and then the $\ker \w_L$ defines an involutive, and hence an integrable,
distribution that is also generated by $\Delta$ and $K$, $K$ being 
the vector field
\be
K={1\over v_z}\G_{\Bbb L}  
\ee
commuting with $\Delta$.
In this way, we can find new local coordinates adapted to the distribution
that allow us to find later on the symplectic form in the quotient manifold, by 
solving the following differential equation systems:
\be \Delta f = 0 \quad Kf= 0.
\ee
\be
\Delta f = 1\quad Kf= 0.
\ee
\be
\Delta f = 0 \quad Kf = 1.
 \ee
Acoording to the solution of the former systems we will do the following
choice for the new coordinates:
\begin{eqnarray}
x_1 &=& {v_x\over v_z}z-x, \quad x_2={v_y\over v_z}z-y, \quad x_3=z \\
y_1 &=& {v_x\over v_z}, \quad y_2={v_y\over v_z}, \quad y_3=\log v_z,
\end{eqnarray}
the inverse change being given by
\begin{eqnarray}
x &=& y_1x_3-x_1, \quad y=y_2x_3-x_2, \quad z=x_3 \\
v_x &=& y_1\exp {y_3}, \quad v_y=y_2\exp {y_3}, \quad v_z=\exp {y_3} .
\end{eqnarray} 
Moreover, in the reduction of the presymplectic form we must point out
that the condition $\Delta n=0$ is written in the new coordinates as 
${\partial n \over \partial y_3}=0$.
Finally, using the change of coordinate former,
 we get after some calculations
the following symplectic form in the quotient manifold
\begin{eqnarray}
\widetilde\w_L &=& d\left[{ny_1\over {\sqrt {y_1^2+y_2^2+1}}} +
{\sqrt {y_1^2+y_2^2+1}}{\partial n\over{\partial y_1}}\right]
\land dx_1  \\ &+& 
d\left[{ny_2\over {\sqrt {y_1^2+y_2^2+1}}} +
{\sqrt {y_1^2+y_2^2+1}}{\partial n\over{\partial y_2}}\right]
\land dx_2
\end{eqnarray}    
which is in full agreement with \cite {Wolfdos}.

As a final comment, let us remark that, even in this case, if we restrict 
ourselves to a region of constant index, we recover the Darboux coordinates
for constant index medium and we can think of the relation betwen the ingoing
and outgoing light rays as a change of Darboux coordinates. Therefore this 
change of Darboux coordinates seems to be again, from an active viewpoint, 
a canonical transformation. The transformations of phase
space will be, in general, non lineal, 
i.e., they generate optical aberrations. It is possible, finally, to analyse
these aberrations using both group theoretical and Lie algebraic tools
(see \cite{Dragt}, \cite{Wolfdos}, and references therein).

\end{document}